\documentclass[pra,aps,twocolumn,showpacs,superscriptaddress]{revtex4}
\usepackage{dcolumn}
\usepackage{bm}
\usepackage{amsmath,amssymb}
\usepackage{graphicx}
\usepackage{hyperref}

\newcommand{\ecos}[1]{\left<\cos^2\theta_{#1}\right>}
\newcommand{\pol}[1]{\alpha^\text{pol}_{#1}}
\newcommand{\ket}[1]{\left|#1\right>}

\begin{document}

\title{Alignment of asymmetric-top molecules using multiple-pulse trains} 

\author{Stefan Pabst}
\affiliation{Argonne National Laboratory, Argonne, Illinois 60439, USA}
\affiliation{Institut f\"ur Theoretische Physik, Universit\"at Erlangen-N\"urnberg, D-91058 Erlangen, Germany}

\author{Robin Santra}
\thanks{Corresponding author.}
\affiliation{Argonne National Laboratory, Argonne, Illinois 60439, USA}
\affiliation{Department of Physics, University of Chicago, Chicago, Illinois 60637, USA}

\date{\today}


\begin{abstract}
We theoretically analyze the effectiveness of multiple-pulse laser alignment methods for asymmetric-top molecules. 
As an example, we choose SO$_2$ and investigate the alignment dynamics induced by two different sequences, each consisting of four identical laser pulses. Each sequence differs only in the time delay between the pulses. Equally spaced pulses matching the alignment revival of the symmetrized SO$_2$ rotor model are exploited in the first sequence. 
The pulse separations in the second sequence are short compared to the rotation dynamics of the molecule and monotonically increase the degree of alignment until the maximum alignment is reached.
We point out the significant differences between the alignment dynamics of SO$_2$ treated as an asymmetric-top and a symmetric-top rotor, respectively. 
We also explain why the fast sequence of laser pulses creates considerably stronger one-dimensional molecular alignment for asymmetric-top molecules. 
In addition, we show that multiple-pulse trains with elliptically polarized pulses do not enhance one-dimensional alignment or create three-dimensional alignment.
\end{abstract} 

\maketitle


Molecular alignment techniques have become important for controlling processes like photo-absorption \cite{BiCl-Science323,PeBu-APL92}, multiphoton ionization \cite{LiLe-PRL90,KuHo-PRL100,PlMc-JPhysB30}, high harmonic generation (HHG) \cite{LaMa-JPhysB33,VeHa-PRL87}, and molecular imaging \cite{ItZe-Nature04,SpDo-PRL04,PaHo-PRA81}.
Alignment of molecules can be achieved with intense laser fields making use of the quadratic Stark effect \cite{StSe-RMP03}.
In general it is true that more intense laser fields create higher degrees of alignment.
However, intense laser fields trigger side effects, like multiphoton ionization and molecular defragmentation, that irreversibly damage molecules \cite{PeNi-JCP127,MaRo-PRA69,LeBl-PRL86}.
For alignment purposes, ionization is an unwanted effect that multiple-pulse alignment techniques try to prevent \cite{CrBu-PRA80}.

Laser alignment can be accomplished adiabatically or impulsively \cite{StSe-RMP03}. In the former case, the laser pulse duration is long compared to the rotational period of the molecule $\tau_\text{rot}$; in the latter case it is short compared to $\tau_\text{rot}$. 
For a given laser intensity, adiabatic alignment leads to a higher degree of alignment than does impulsive alignment; however, enhancing alignment through several consecutive, non-overlapping laser pulses is only possible in the impulsive regime.
Theoretical and experimental studies with up to three laser pulses, where pulse separations, pulse intensities, and pulse shapes were systematically varied, have been performed \cite{LeAv-PRL90,BiPo-PRL92,GuRo-PRA77}.
Recently, field-free alignment of N$_2$ was reported in an experiment with eight identical, Fourier transform limited, consecutive laser pulses \cite{CrBu-PRA80}. All eight pulses were separated by the rotational period $\tau_\text{rot}=1/(2B)$ \cite{CrBu-PRA80,BuSa-JCP129}. 
The degree of alignment achieved in Ref. \cite{CrBu-PRA80} with eight pulses is much greater than the alignment induced by a single ionization-limited pulse.

Attempts to use a sequence of pulses to enhance alignment have so far been focused on linear or symmetric-top molecules, which possess well-defined alignment revivals separated by $\tau_\text{rot}$.
The irregular or incommensurable spacings of the rotational energy levels for asymmetric-top molecules prevent full rephasing of the rotational wave packet \cite{RoGu-PRA73} and, therefore, the appearance of periodic alignment revivals. 
Experiments involving asymmetric-top molecules with one-dimensional, field-free alignment \cite{PePo-PRL91,PePo-PRA70,PoPe-JCP121,RoGu-PRA73,HoVi-PRA75} and three-dimensional alignment using two linearly polarized laser pulses \cite{UnSu-PRL94,LeVi-PRL97} have been reported. 

Another source of rotational wave packet dephasing is centrifugal distortion, which becomes relevant when rotational states with high angular momentum, needed to get high degrees of alignment, are populated. This effect is not limited to asymmetric-top molecules and affects linear and symmetric-top molecules as well \cite{BrCo-PRA78}. 

In this work, we extend the idea of multiple-pulse alignment to rigid, asymmetric-top molecules, omitting the effect of additional dephasing through centrifugal distortion. 
Specifically, we theoretically investigate the feasibility of enhancing one-dimensional alignment.
We consider two different pulse trains, each consisting of four identical laser pulses. 
The pulses are equally separated by the revival period in the first pulse train. This strategy follows Ref. \cite{CrBu-PRA80}. 
In the second train, pulses are separated such that the molecule experiences an additional kick when it reaches the maximum alignment induced by the previous pulse, resulting in a monotonic increase in the degree of alignment \cite{AvAr-PRL87}. 
Furthermore, we point out the consequences of approximating an asymmetric-top rotor as a symmetric-top rotor.
We then investigate 4-pulse trains using elliptically polarized laser pulses and ask the questions whether one-dimensional alignment is enhanced in comparison to the use of linearly polarized pulses and if it is possible to create field-free three-dimensional alignment.

We do not review our numerical propagation method, which is described in Ref. \cite{PaHo-PRA81}.
In the following, we will choose to subject the molecule SO$_2$ to an electric laser field,
\begin{eqnarray}
	\label{eq.1}
	{\vec E}(t)
	&=&
	\sqrt{8\pi\,I(t)/c}
	\bigg[
		\epsilon_x\,\vec e_x \cos(\omega t)
		+
		\epsilon_z\,\vec e_z \sin(\omega t)  
	\bigg],
\end{eqnarray}
where $c$ is the speed of light, $I(t)$ is the cycle-averaged laser intensity, and $\epsilon_x,\epsilon_z$ are the minor and major field components with $\epsilon_x^2\leq\epsilon_z^2$ and $\epsilon_x^2+\epsilon_z^2=1$. We set $\epsilon_x=0$ to describe linearly polarized light.
The molecule SO$_2$ has the rotational constants $A=0.3442$~cm$^{-1}$, $B=0.2935$~cm$^{-1}$, $C=2.028$~cm$^{-1}$ \cite{FrBe_JRaSpec31,Herzberg-book3,comm1} and
polarizabilities $\pol{aa}=20.80~a_0^3$, $\pol{bb}=18.66~a_0^3$, $\pol{cc}=31.32~a_0^3$ \cite{XeMa-CPL319}, where $a_0$ denotes the Bohr radius. The nuclear spin statistical weights of SO$_2$ are 1 if $\ket{J\tau M}\in A,B_a$ and 0 if $\ket{J\tau M}\in B_c,B_b$, where $\ket{J\tau M}$ denotes a rotational eigenstate of an asymmetric-top rotor and $B_a,B_b,B_c$ and $A$ are irreducible representations of $D_{2}$ (isomorphic to $C_{2v}$) \cite{Bunker-book}. 

\begin{figure}[ht]
  \begin{center}
    \includegraphics[clip,width=\linewidth]{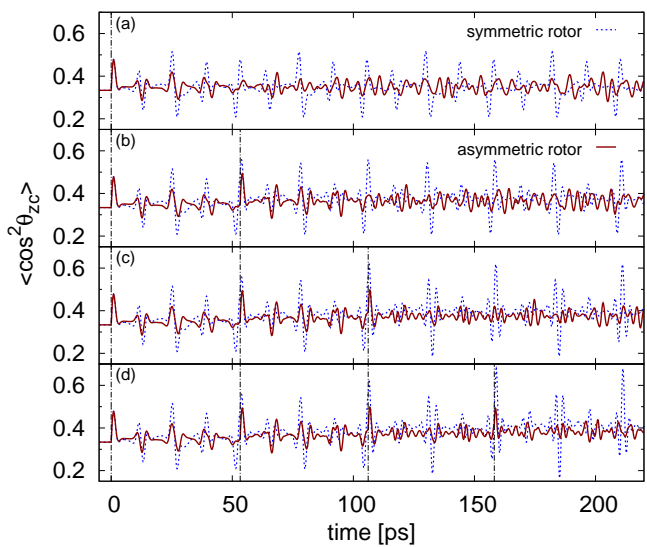}
    \caption{(Color online) Alignment dynamics of SO$_2$ treated, respectively, as a symmetric-top (dashed) and an asymmetric-top rotor (solid), at rotational temperature $T=10$~K.
    From panel (a) to (d) the number of pulses successively increases from one to four.
    The linearly polarized laser pulses, which are indicated by vertical dashed lines, have a peak intensity of $20$~TW/cm$^2$ and a pulse duration of $50$~fs.}
    \label{fig.1}
  \end{center}
\end{figure}

Figure \ref{fig.1} presents the alignment dynamics of SO$_2$, treated as a symmetric-top and an asymmetric-top rotor, respectively, for a sequence of up to four linearly polarized, consecutive Gaussian-shaped laser pulses spaced equally by $\tau_\text{rot}=1/(A+B)$. 
The laser intensity is 20~TW/cm$^2$ and the pulse duration (FWHM) is $50$~fs. 
The chosen rotational temperature of 10~K is a realistic estimate that has been experimentally achieved for SO$_2$ \cite{LeVi-PRL97}.
The symmetric-top rotor is approximated by symmetrization of the $a$ and $b$ axes, i.e.,
$A,B \rightarrow (A+B)/2$ and $\pol{aa},\pol{bb} \rightarrow (\pol{aa}+\pol{bb})/2$.

In the symmetric-top model, the molecules show the expected revival dynamics in Fig. \ref{fig.1}. 
By increasing the number of laser pulses (Fig. \ref{fig.1}(a)-(d)), the maximum alignment increases monotonically from $\ecos{zc}=0.52$ (one pulse) to $\ecos{zc}=0.68$ (four pulses).
When SO$_2$ is treated exactly as an asymmetric-top rotor, no regularly repeating alignment motion can be identified.
The dephasing, due to the incommensurable spacing between the rotational energy levels, increases with time and is the reason why the maximum alignment achieved after the fourth laser pulse is weaker than the alignment created directly after the third laser pulse (cf. Fig. \ref{fig.1}d).

\begin{figure}[ht]
  \begin{center}
    \includegraphics[clip,width=\linewidth]{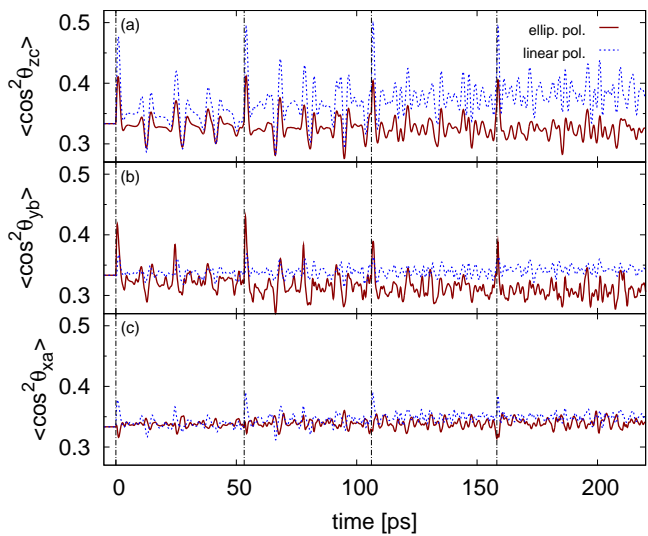}
    \caption{(Color online) Alignment dynamics of SO$_2$ for linearly polarized (dashed) and elliptically polarized (solid) laser pulses with $\epsilon^2_z=0.5462$. The peak intensity associated with the $z$ direction (20~TW/cm$^2$) is kept the same in both cases, as is the pulse duration of 50~fs. SO$_2$ is treated in both cases as an asymmetric-top rotor.}
    \label{fig.2}
  \end{center}
\end{figure}

Another question we want to address is whether multiple elliptically polarized laser pulses can be used to create three-dimensional alignment in the same manner as linearly polarized laser pulses can be used to achieve one-dimensional alignment.
Figure \ref{fig.2} shows a direct comparison of the alignment of SO$_2$ (treated as an asymmetric-top rotor) for four linearly ($\epsilon_x=0)$ and four elliptically polarized laser pulses ($\epsilon^2_z=0.5462$). The specific value of $\epsilon^2_z$ for elliptically polarized light is chosen such that optimal three-dimensional alignment is obtained \cite{RoGu-PRA77}.
The laser intensity associated with the $z$ direction is the same for both types of polarized laser pulses. The total intensity of the elliptically polarized laser pulses is adjusted accordingly.

The additional electric field in the perpendicular $x$ direction decreases the $\ecos{zc}$ alignment (cf. Fig. \ref{fig.2}(a)) and simultaneously increases the alignment of the body-fixed $c$ axis in the $x$ direction ($\ecos{xc}$).
However, elliptically polarized laser pulses improve the alignment of the molecules in the elliptical polarization plane of the laser pulses ($zx$ plane), which is given by $\ecos{yb}$ (see Fig. \ref{fig.2}(b)).
The alignment dynamics of $\ecos{xa}$ are counterintuitive and show an antialignment effect rather than an alignment effect (cf. Fig. \ref{fig.2}(c)). 
By analyzing all $\ecos{lm}$, we find that each molecular axis is aligned or antialigned simultaneously in the $x$ and $z$ directions. From the relation, $\sum_{m'}\ecos{lm'}=\sum_{l'}\ecos{l'm}=1\ \forall\, l,m$, it follows that the alignment in the $y$ direction is reversed from the alignment in the $x$ and $z$ directions. 
Since the molecular alignments $\ecos{zc}$ and $\ecos{xc}$ are strongly pronounced, the molecular $a$ and $b$ axes are antialigned in these two space-fixed directions and, hence, aligned in the $y$ direction ($\ecos{ya}$ and $\ecos{yb}$).
Only in the adiabatic limit would we see strong alignment in $\ecos{xa},\ecos{yb}$ and $\ecos{zc}$.

\begin{figure}[ht]
  \begin{center}
    \includegraphics[clip,width=\linewidth]{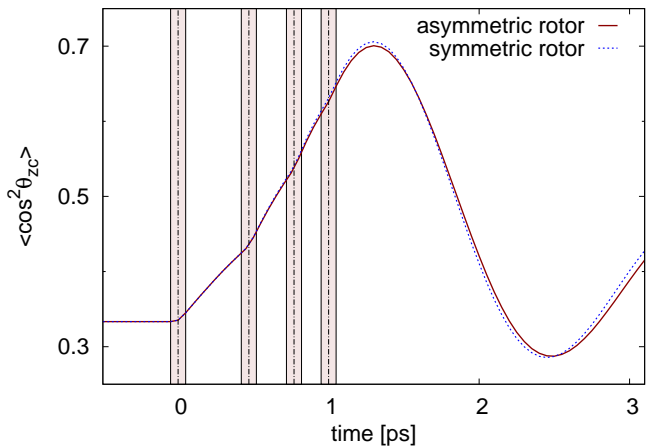}
    \caption{(Color online) Alignment dynamics of SO$_2$ treated, respectively, as a symmetric-top (dashed) and an asymmetric-top rotor (solid). The four laser pulses are linearly polarized with a peak intensity of 20~TW/cm$^2$. The shaded areas indicate the FWHM-width of 50~fs pulses centered around the vertical, dashed lines.}
    \label{fig.3}
  \end{center}
\end{figure}

Improving the degree of alignment of symmetric-top molecules by applying consecutive laser pulses at maximum alignment can be done either at the first alignment peak directly after the previous pulse or at later times at alignment revivals.
In the case of asymmetric-top molecules, the accessibility of revivals is limited to the very first revivals (cf. Fig. \ref{fig.1}) and the maximum achievable degree of alignment is reduced in comparison to the linear and symmetric-top rotor models.
However, the dynamics immediately following the first laser pulse are almost identical for both rotor models, since the dephasing effects are still small.
 It is in this time frame, where the very first alignment peak occurs.
Therefore, applying subsequent laser pulses close to the very first laser pulse promises better alignment. In that way the alignment is increasing monotonically till it has reached its maximum degree of alignment.
In Fig. \ref{fig.3} such a pulse sequence is presented. The alignment profiles for both rotor models are almost identical with a maximum alignment comparable with the revival kicking technique for symmetric-top rotors shown in Fig. \ref{fig.1}(d).
The alignment response, which is the time after a pulse until maximum alignment is reached, decreases with the number of pulses---and so does the spacing between neighboring laser pulses \cite{AvAr-PRL87}. This limits the maximum number of laser pulses that may be employed to accomplish field-free alignment. 
However, the maximum degree of alignment for both rotor models differs by less than 1\% in Fig. \ref{fig.3}.

\begin{figure}[hb]
  \begin{center}
    \includegraphics[clip,width=\linewidth]{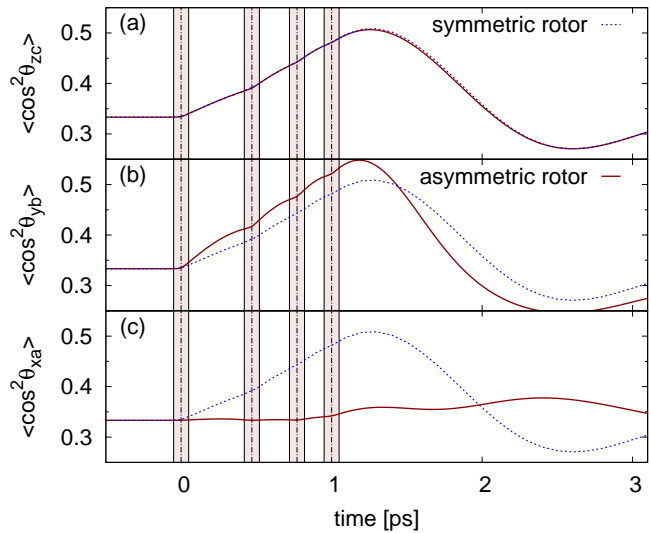}
    \caption{(Color online) Alignment dynamics of SO$_2$ treated, respectively, as a symmetric-top (dashed) and an asymmetric-top rotor (solid). The four laser pulses are elliptically polarized with $\epsilon^2_z=0.5462$, a peak intensity of 36.6~TW/cm$^2$ ($I_z=20$~TW/cm$^2$), and a pulse duration of 50~fs.}
    \label{fig.4}
  \end{center}
\end{figure}

By exploiting the same method for elliptically polarized laser pulses, we find that the alignment dynamics for the symmetric-top and asymmetric-top rotor models are identical in terms of $\ecos{zc}$ (cf. Fig. \ref{fig.4}(a)), as in the linearly polarized case (cf. Fig. \ref{fig.3}). 
The alignment of the asymmetric-top rotor model in the polarization plane, which is characterized by $\ecos{yb}$, is enhanced compared to the symmetric-top rotor model (cf. Fig. \ref{fig.4}(b)).
During all four laser pulses the alignment $\ecos{xa}$ stays almost isotropic and increases only slightly. 
However, $\ecos{xa}$ does not show any antialignment within the first few picoseconds (see Fig. \ref{fig.4}(c)) as Fig. \ref{fig.2}(c) shows for the revival-kicking pulse sequence. 
The lack of $\ecos{xa}$ alignment is not a problem of intensity; it is the result of the rich rotational dynamics of the asymmetric-top rotor SO$_2$. Only in the limit of adiabatic alignment does this motion cease and all three molecular axes become well-aligned. 

In conclusion, we studied multiple-pulse alignment of asymmetric-top molecules, using SO$_2$ as an example.
We showed that approximating an asymmetric-top molecule as a symmetric-top rotor has significant consequences for the alignment dynamics; specifically, alignment revivals do not occur for asymmetric-top molecules.
The dephasing of the rotational wave packet for asymmetric-top molecules limits the effectiveness of aligning the molecules by multiple pulses applied at alignment revivals. 
Enhanced alignment for asymmetric-top molecules can be better accomplished by a fast train of pulses. Here the time delays between consecutive pulses are small compared to the rotational time scale such that dephasing effects are minimized. 
Therefore, for this method the maximum degree of alignment is not affected by the more complex rotational dynamics of an asymmetric-top molecule. 
However, when elliptically polarized pulses are used, none of these approaches attains significant three-dimensional alignment or improves the one-dimensional alignment further.
We conclude that a train of elliptically polarized laser pulses is not suitable for achieving field-free three-dimensional alignment.

\acknowledgments
We thank James P. Cryan, Christian Buth, and Ryan N. Coffee for inspiring discussions, and Cassandra Hunt for comments on the manuscript.
This work was supported by the Office of Basic Energy Sciences, U.S. Department of Energy under Contract No. DE-AC02-06CH11357.


\begin{thebibliography}{999}
\bibitem{comm1} In contrast to spectroscopic convention, the rotational constants $A,B$ and $C$ are not ordered according to magnitude. We reorder the rotational constants such that for perfect alignment the molecular frame ($a,b,c$) coincides with the space-fixed frame ($x,y,z$).
\end{thebibliography}
\end{document}